\documentclass{article}
\usepackage{spconf,amsmath,graphicx}
\usepackage{amssymb}
\usepackage{stfloats}
\usepackage{multirow}
\usepackage{booktabs}
\usepackage{makecell}
\usepackage{flushend}
\usepackage{cite}

\title{TDCGAN: Temporal Dilated Convolutional Generative Adversarial Network for End-to-end Speech Enhancement}
\name{Shuaishuai Ye, Xinhui Hu and Xinkang Xu}
\address{Hithink RoyalFlush AI Research Institute, Zhejiang, China}
\email{\{yeshuaishuai,huxinhui,xuxinkang\}@myhexin.com}
\begin{document}
\ninept
\maketitle
\begin{abstract}
In this paper, in order to further deal with the performance degradation caused by ignoring the phase information in conventional speech enhancement systems, we proposed a temporal dilated convolutional generative adversarial network (TDCGAN) in the end-to-end based speech enhancement architecture. For the first time, we introduced the temporal dilated convolutional network with depthwise separable convolutions into the GAN structure so that the receptive field can be greatly increased without increasing the number of parameters. We also first explored the effect of signal-to-noise ratio (SNR) penalty item as regularization of the loss function of generator on improving the SNR of enhanced speech. The experimental results demonstrated that our proposed method outperformed the state-of-the-art end-to-end GAN-based speech enhancement. Moreover, compared with previous GAN-based methods, the proposed TDCGAN could greatly decreased the number of parameters. As expected, the work also demonstrated that the SNR penalty item as regularization was more effective than $L1$ on improving the SNR of enhanced speech.
\end{abstract}
\section{Introduction}
\label{sec:intro}

Speech enhancement (SE) is an indispensable front-end module in intelligent speech devices\cite{HearAid} and speech applications, such as automatic speech recognition (ASR)\cite{ASR}. One of its important functions is to enhance the robustness of speech back-end systems in complicated scenarios, so as to ultimately improve the practicalities of these devices and applications. In order to improve SE performance in complicated scenarios, plenty of SE methods based on deep machine learning have been proposed during recent years\cite{xuyong-ARA,Tan2018A,URTF,FCRN,TFNLM,TFLLSC}. 

However, most of the previous methods operate on spectrum characteristics, (e.g. power spectrum and magnitude spectrum) by calculating the short-time Fourier transform (STFT) and inverse STFT (iSTFT). Such operations seriously weaken perceptual quality of speech because they ignore the phase information of speech signals\cite{Tan2018A,URTF,FCRN,TFNLM,TFLLSC}. So, the spectrum characteristics are always regarded as non-optimal representations for SE tasks. To make full use of the phase information of speech signals, the end-to-end methods, in which the enhancements are performed in a waveform-to-waveform manner, are actively studied\cite{lu2013speech,E2EWaveNet,E2ESSE2018,SEGAN2017}. Compared with traditional methods for SE, the end-to-end SE directly operate on raw waveform with both magnitude and phase information by replacing STFT (in encoder) and iSTFT (in decoder) by neural networks, which eliminates the processing of speech signal from time domain to frequency domain so as to reduce the computational complexity\cite{E2ETCNN}. 

As a novel generative network architecture, generative adversarial network (GAN)\cite{GAN}, composed of generator and a discriminator, have attracted many researchers' attention in recent years. Compared with conventional neural networks, GAN allows to model more complex tasks and generate higher-quality samples by an adversarial training manner\cite{GAN}. In SE task, GANs have also achieved great successes  \cite{SEGAN2017,SEWGAN2018,SERGAN2019,CPGAN2020}. The pioneering work of speech enhancement GAN (SEGAN)\cite{SEGAN2017} opened the way for GAN-based end-to-end SE, and it produces less speech distortion and removes noise more effectively than traditional methods do\cite{SEGAN2017}. Since then, various variants of GAN have been applied to end-to-end SE tasks, such as Wasserstein GAN\cite{SEWGAN2018}, relativistic GAN\cite{SERGAN2019} and context pyramid GAN\cite{CPGAN2020}, all of them have made large achievements for SE tasks. 

Nevertheless, most of prior GAN-based end-to-end SE methods basically followed the architecture of the above SEGAN\cite{SEWGAN2018,SERGAN2019}. The architecture has following shortcomings: (1) It simply uses standard convolutional networks, so it does not make full use of context information to better predict current enhanced samples. (2) It has a relatively large number of parameters. So, the enhanced speech quality and intelligibility of these GAN-based SE models is relatively low, and the large model complexity makes it relatively difficult to train. To address these shortcomings, we proposed an end-to-end time-domain SE system with a temporal dilated convolutional generative adversarial network (TDCGAN). In this system, a temporal dilated convolutional network (TDCN) with depthwise separable convolutional network (DSCN)\cite{FCNN} was introduced to the GAN.The TDCN with DSCN increases the receptive field of network by a substantial amount while the number of parameters  will not be increased. So, such mechanism is able to improve network's modeling capability. This architecture has succeeded greatly in the fields of time-domain audio separation\cite{TasNet3,TasNet2}, sequence learning\cite{sequence} and action recognition\cite{actionreg}. In the proposed method, the generator of the GAN, which is composed of an encoder, a temporal dilated convolutional mask estimator (TDCME) and a decoder, is used to estimate clean speech signals from  noisy speech signals, while the discriminator of the GAN, which is similar to the architecture of SEGAN's, is used for guiding training of the generator by calculating some distance, such as Wasserstein distance\cite{IWGAN} between clean and enhanced (generative) speech distributions.

Our contributions are embodied in following aspects: (1) We proposed a TDCGAN architecture with TDCN and DSCN which were effective in other area such as audio separation, for the time-domain end-to-end SE tasks. (2) For the first time, we explored the effect of signal-to-noise ratio (SNR) penalty item as regularization of the loss function of generator on improving the SNR of enhanced speech signals and verified its effectiveness. (3) With above processes, the performances of SE were improved, while the trainable parameters were greatly reduced. 

 \begin{figure}[t]
  \centering
  \includegraphics[width=\linewidth]{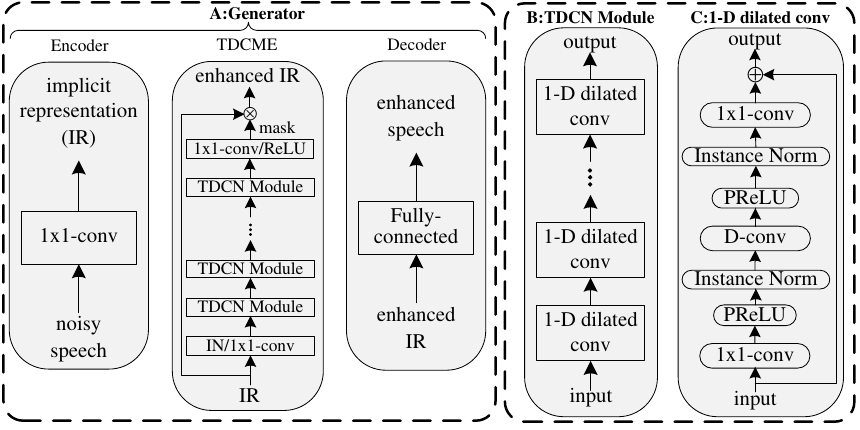}
  \caption{(A): the block diagram of the TDCGAN system. IN/1$\times$1-conv represents preforming instance normalization on input features before 1$\times$1 convolutions. ReLU is a nonlinear activation function. (B): temporal dilated convolutional network module. (C): 1-D dilated convolutional block with residual network. PReLU is also nonlinear activation function.}
  \label{fig:TGCGAN-generator}
\end{figure}

\section{Speech Enhancement Using GAN}
\label{sec:speech}

Within a GAN, the generator is used for mapping some prior distribution $\mathcal{Z}$ to another distribution $\mathcal{X}$.With this mapping, we expect to fool discriminator. The discriminator's main task is either to distinguish the fake (generative) distribution from the true distribution or to compute the distance metrics, such as Wasserstein distance between the fake and true distributions. The generator and discriminator continuously optimize alternately themselves at an equilibrium.

In the mentioned work of the SEGAN, the generator is regarded as a denoiser to preform the mapping from noisy speech to clean speech, meanwhile the discriminator is used as binary classifier to distinguish clean speech from enhanced speech. The generator of SEGAN is structured similarly to an auto-encoder with one-dimensional fully-convolutional networks and skip-connections. The discriminator of SEGAN follows the same structure as generator’s encoder. With such a network architecture, the SEGAN produces less speech distortion and removes noise more effectively than traditional methods do. However, the SEGAN's performances and its model size are still found having space for improvements.

Currently, the mainstream GANs for speech enhancement include least-square GAN (LSGAN)\cite{LSGAN}, relativistic GAN (RGAN)\cite{RGAN} and Wasserstein GAN with gradient penalty item (WGAN-GP)\cite{IWGAN}. 

LSGAN and RGAN improve the quality of the generative samples, though they don't solve the main problems, such as unstable training and difficult converging. As a stable version of the family, WGAN-GP can more clearly guide training of the model by optimizing the Wasserstein distance between clean and enhanced speech distribution than LSGAN and RGAN. The loss functions of the WGAN-GP are as follows.
%
\begin{equation}
\mathcal{L}_D = -\mathbb{E}_{{x_\mathtt{rp}}}\left[C(x,y)\right]+\mathbb{E}_{{x_\mathtt{fp}}}\left[C(\hat{x},y)\right]+\lambda_{gp}\mathcal{L}_{gp}
\end{equation}
\begin{equation}
\mathcal{L}_G = -\mathbb{E}_{{x_{\mathtt{fp}}}}\left[C(\hat{x},y)\right]
\end{equation}
In these formulas, the $x$, $\hat{x}$, $y$ are the clean, enhanced (generative) and noisy speech signals respectively. $\hat{x}\triangleq{G(y)}$, $x_{\mathtt{rp}}\triangleq{(x,y)}\sim{p(x,y)}$ which represents the joint probability distribution of the $x$ and $y$, and $x_{\mathtt{fp}}\triangleq{(\hat{x},y)}\sim{p(\hat{x},y)}$. The $C$ is discriminator, the $G$ is the generator, $\mathcal{L}_{gp}$ is the gradient penalty item and $\lambda_{gp}$ controls the magnitude of the $\mathcal{L}_{gp}$.

 It has been proven that the WGAN-GP with simplified zero-centered gradient penalty can locally converge under suitable assumptions\cite{R1R2}. Due to such characteristics, we basically selected the WGAN-GP to follow for our speech enhancement task. Similar to SEGAN\cite{SEGAN2017}, in order to minimize the distance between generative and clean examples, we add $L1$ regularization to the loss function of generator. The magnitude of the $L1$ is controlled by a hyper-parameter $\lambda_{L1}$.Therefore, the loss function of the generator becomes  $\mathcal{L}_G+\lambda_{L1}\Vert{\hat{x}-x}\Vert_1$.

\section{Our Proposed Model : TDCGAN}
\label{sec:tdcgan}

 SE task is used to separate clean speech signals and noise signals. Audio separation task is used to separate speech signals of different audio sources such as different speakers. In essence, SE task is very similar to audio separation task. So, in this study, we introduced the excellent time-domain audio separation network (TasNet)\cite{TasNet3} to an architecture of GAN and proposed a temporal dilated convolutional generative adversarial network (TDCGAN) for speech enhancement. Within this new type of GAN architecture, the generator is based on temporal dilated convolutional networks (TDCN) with non-causal convolutions\cite{E2EWaveNet,TasNet3,TasNet2} and depthwise separable convolutions (DSC)\cite{FCNN}. And the discriminator is based on convolutional neural network with DSC. 
 Compared with the discriminator of SEGAN, the TDCGAN's uses fewer convolutional layers and replaces standard convolutions with DSCs. As advantages of the TDCGAN, its TDCN with non-causal convolutions can allow the network to model long-term dependencies of speech signal\cite{actionreg}, and its DSC can reduce the number of trainable parameters\cite{FCNN}.

\subsection{Generator}
\label{ssec:generaor}

The generator within the TDCGAN is equivalent to a denoiser. Its architecture is depicted in Figure \ref{fig:TGCGAN-generator} (A). The generator uses the same structure as the TasNet\cite{TasNet3}, but with a few differences: (1) it replaces the last convolutional layer with a fully-connected layer to generate more realistic speech samples. (2) it uses instance normalization (IN)\cite{IN} instead of channel normalization or global normalization. The generator is composed of an encoder, a temporal dilated convolutional mask estimator (TDCME) module, and a decoder. The encoder with one-layer convolution is used to extract implicit representation (IR) for noisy speech signal. The TDCME is used to extract mask to enhance noisy IR. It consists of an input convolutional layer, N\(\in \mathbb{N_+}\) stacked TDCN module(s) and output convolutional layer followed by a nonlinear activation function rectified linear unit (ReLU). The enhanced speech signal is then reconstructed by the enhanced IR using a decoder module with one fully-connected layer. The key features of the generator are presented below:


\begin{figure}[t]
  \centering
  \includegraphics[width=\linewidth]{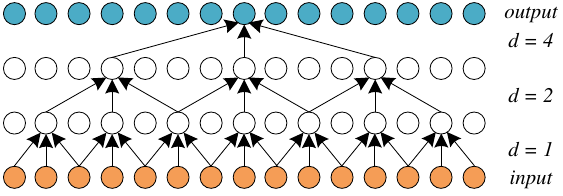}
  \caption{An example of non-causal dilated convolution with kernel of size 3.}
  \label{fig:TGCGAN-non}
\vspace{-1.5em}  
\end{figure}

\textbf{Non-causal, dilated, depthwise separable convolutions}. The generator based on TasNet makes use of non-causal, dilated, depthwise separable convolutions\cite{TasNet3,TasNet2}. The dilated convolutions with dilation factors allow the network to expand the receptive field in each layer and make full use of more speech information to enhance accurately current sample. Similarly, in order to further expand receptive field to future samples within  afforded latency in model response, we add the non-causality to convolutions, as shown in every layer of the Figure \ref{fig:TGCGAN-non}, to employ future samples to enhance current samples very well. The DSC consists of  a \textit{depthwise} convolutional network and a \textit{pointwise} convolutional network. Compared with standard convolutional network, the DSC greatly decreases the number of parameters without performance degradation. The DSC has been proven to be effective in speech separation and machine translation\cite{FCNN}.

\textbf{TDCN module}. TDCN module uses  M \(\in\mathbb{N_+}\)  stacked 1-D dilated convolution block(s) with exponentially increasing dilation factors $1, 2, ... , 2^{M-1}$, as shown in Figure \ref{fig:TGCGAN-generator} (B).  The dilation factors increase exponentially to ensure a sufficiently large temporal context window to make use of the long-term dependencies of the speech signal, as shown in Figure \ref{fig:TGCGAN-non}.

\textbf{1-D dilated convolution block}. The 1-D dilated convolution block is shown in Figure \ref{fig:TGCGAN-generator} (C). The block is composed of three parts, namely input-end of 1$\times$1 convolutional layer (ICL), depthwise separable convolutional layer (DSCL) and output-end of 1$\times$1 convolutional layer (OCL). The ICL and DSCL are all followed by an IN and a nonlinear activation function PReLU. The output of the block is a residual between the original input and the ouput of OCL. In the 1-D convolution block, residual path ensures that the gradients can be transferred in a quite deep network and the problem of gradient vanishing can be improved.

\textbf{Signal-to-noise ratio penalty item}. In the generator of a GAN for SE task, the additional loss penalty item facilitates guiding model to converge and generating more realistic samples\cite{SEGAN2017,SEWGAN2018,SERGAN2019,CPGAN2020}. Motivated by the good SE performance for speech recognition\cite{ASR}, we explore using signal-to-noise ratio (SNR) to replace the regularization $L1$ for the proposed GAN. The SNR is formulated as follows:
\begin{equation}
  \mathcal{L}_{SNR} = -10log(\frac{\Vert{x}\Vert^2}{\Vert{x-\widehat{x}}\Vert^2})
  \label{eq10}
\end{equation}

Where ${\Vert\Vert}^2$ is the $L2$ norm. The magnitude of the SNR is controlled by a hyper-parameter $\lambda_{SNR}$. So, the loss function of the generator becomes  $\mathcal{L}_G+\lambda_{SNR}\mathcal{L}_{SNR}$. Our preliminary experiments showed that the SNR penalty item was more effective than $L1$ on improving the SNR of speech signals. 

\textbf{Instance normalization}. In most GAN-based SE methods, batch normalization (BN) is quite effective\cite{SEGAN2017,SEWGAN2018}. However, as a domain adaptive normalization with learning the domain mean and variance, BN is more suitable for discriminative model than for generative model. More importantly, the performance of BN is not stable with the change of batch size. Therefore, to address these problems, we introduce the instance normalization (IN)\cite{IN} to the generator of the proposed GAN. Unlike to BN, IN performs normalization on every speech feature map in single instance of every batch, so it can not only accelerate the convergence of  the model, but also ensure the independence among speech features\cite{IN}.

\subsection{Discriminator}
\label{ssec:discriminator}

For the discriminator of the TDCGAN, we adopted a similar architecture to the SEGAN's\cite{SEGAN2017}, with two differences: (1) the last nonlinear activation function sigmoid is removed, (2) the standard convolutions are replaced with the depthwise separable convolutions to decrease the number of trainable parameters. The numbers of kernels in every layer are 16, 32, 32, 64, 128, 128, 256, 512 and 1024 respectively.

To deal with the problem of unstable training process existed in the original GANs , we introduced zero-centered gradient penalties\cite{R1R2} to our discriminator, because it has been proven to facilitate GAN's training to locally converge. There are two versions for the zero-centered gradient penalties: one is on real data and another is on fake data. The regularization terms corresponding to them are formulated as follows\cite{R1R2}.
\begin{equation}
  R_1(\psi) =\frac{\gamma}{2}E_{p_\mathtt{r}(x)} \left[\Vert{\bigtriangledown{D_\psi}(x)}\Vert^2\right]
  \label{eq2}
\end{equation}
\begin{equation}
  R_2(\phi,\psi) =\frac{\gamma}{2}E_{p_\mathtt{f}(x)}\left[\Vert{\bigtriangledown{D_\psi(x)}}\Vert^2\right]
  \label{eq3}
\end{equation}
where $\psi$ and $\phi$ are the variables of discriminator and generator respectively, $p_\mathtt{r}(x)$ denotes real data distribution, $p_\mathtt{f}(x)$ denotes fake data distribution, $\bigtriangledown$ is the sign of computing gradient, and $\gamma$ is a regularization parameter.

In the work, we added both $R_1$ and $R_2$ to the discriminator's loss function. The magnitudes of $R_1$ and $R_2$  are controlled by regularization parameter $\gamma$. Therefore, the loss function of discriminator finally becomes $\mathcal{L}_D+(R_1+R_2)$.

\section{Experiments}
\label{sec:experiments}

\subsection{Dataset}
\label{ssec:dataset}

The experiments were conducted on a simulation database\footnote{https://datashare.is.ed.ac.uk/handle/10283/1942} which is generated from two open data sources : speech data supplied by the Voice Bank corpus\cite{Voice} and environmental sounds provided by the Diverse Environments Multichannel Acoustic Noise Database (DEMAND)\cite{DEMAND}. The speech dataset were downsampled from 48KHz to 16KHz for our experiments. The dataset contains 12396 utterances recorded by 30 speakers, 28 (11572 utterances) of which are used as training set and 2 (824 utterances) are used as test set. The training set are corrupted with 10 types of noise at four SNR levels (0 dB, 5 dB, 10 dB and 15 dB) to build a multi-noise types and multi-SNR conditions training set. The test set is corrupted with 5 types of unseen noise with 4 SNR levels (2.5dB, 7.5dB, 12.5dB and 17.5dB).

\subsection{Experimental setups}
\label{ssec:setups}

We divided speech into frames by sliding the window with frame length of 16384 samples and frame shift of 8192 samples. During the test stage, similar to SERGAN\cite{SERGAN2019} and CP-GAN\cite{CPGAN2020}, we concatenated the enhanced speech segments with frame shift of 8192 samples by averaging the corresponding overlapping samples. we applied a high-frequency pre-emphasis filter of coefficient 0.95 to all input samples during training stages and testing stages, and the output was correspondingly de-emphasized during testing stages.

The model was trained using Adam optimizer for 100 epochs with a batch size of 16. In order to reduce training time, we apply two-timescale-update-rule\cite{TTUR} with different the learning rates of 3$\times{10^{-4}}$ for discriminator and 2$\times{10^{-4}}$  for generator. In addition, we set the weight factor $\lambda_{SNR}$ and $\lambda_{L1}$ to 10 and 100 respectively based on our preliminary experiments and regularization parameter $\gamma$ of $R1$ and $R2$  to 10 according to the experimental results of \cite{R1R2} .


\begin{table}[th]
  \renewcommand\arraystretch{1.5}
  \caption{The detailed network parameters of the generator. B is batch size. Y$_n$ represents 1-D dilated convolution block, where n=1, 2, $\cdots$, 8. The kn, ks, and df are the numbers of kernel, kernel size and dilated factor respectively.}
  \setlength{\tabcolsep}{0.05mm}{
  \label{tab:parameters-list}
  \centering
  \scriptsize
  \begin{tabular}{|c|c|c|c|c|}
    \hline
    {Module} & { Components}&{(kn, ks, df)} & {Input-size} & {Output-size} \\
    \hline
    {Encoder} & $1\times1-conv$ &  (512, 32, 1) & B$\times$16384 & B$\times$512$\times$1023 \\\cline{1-5}
   
    \multirow{12}{*}{TDCME }
      & $(IN)/1\times1-conv$ & (128, 1, 1)
      & B$\times$512$\times$1023  
      & B$\times$128$\times$1023 \\\cline{2-5}
  
      &\makecell[l]{ \\
                   \ \ TDCME = 4\ $\times$\ {TDCN} \\ 
                   \\
                   \ \ TDCN  \   = ($Y_1, Y_2, \cdots, Y_8$) \\
                   \\
                   $
                   \ \ {Y_n} = \left\{ 
                                       \begin{array}{l}
                                            1\times1-conv \\ \\
                                            {IN/PReLU} \\ \\
                                            {D-conv} \\ \\
                                            {IN/PReLU} \\ \\
                                            1\times1-conv \\
                                       \end{array}
                                \right.
                                $
              \\ \ }

      &\makecell[c]{\\ \\ \\ \\ (128, 1, 1) \\ \\ \\ \\ (512, 3, 2$^{n-1}$) \\ \\ \\ \\ (128, 1, 1)}

      & B$\times$128$\times$1023
      & B$\times$128$\times$1023 \\\cline{2-5}
    
      & $1\times1-conv$ & (512, 1, 1) & B$\times$128$\times$1023 & B$\times$128$\times$1023 \\\cline{1-5}
   
    Decoder  & $fully-connected$ &$-$& B$\times$128$\times$1023 & B$\times$16384 \\
    \hline
  \end{tabular}}
\vspace{-1.0em}
\end{table}

The detailed network parameters of the generator are summarized as shown in Table \ref{tab:parameters-list}.

In the discriminator of the TDCGAN, 9 depthwise separable convolutions, one 1$\times$1 convolution  and one fully-connected layer are employed to compute the Wasserstein distance between clean speech and enhanced distributions. The number of kernels of  9 depthwise separable convolutions with kernel size 3 and stride 2 are 16, 32, 32, 64, 128, 128, 256, 512 and 1024 respectively, and the number of kernel of the 1$\times$1 convolution with kernel size 1 and stride 1 is 1.

\subsection{Evaluation metrics and baselines}
\label{ssec:evalution}

In this work, we adopted following six evaluation metrics to evaluate SE performances : PESQ with range of [-0.5, 4.5] and STOI for [0, 1], segSNR for [0, +$\infty$] Csing for [1, 5], Cbak for [1, 5], Covl for [1, 5]\cite{PESQ,STOI,segSNR}. All metrics (the higher the better) compare the enhanced signal with the clean reference of the 824 test set files. These metrics are computed by using a public code set
\footnote{https://www.crcpress.com/downloads/K14513/K14513\_CD\_Files.zip}.

We compared our proposed method with other 3 GAN-based baseline methods for which the identical dataset was employed. The baselines included the SEGAN\cite{SEGAN2017} which was the pioneer work for GAN-based SE, the SERGAN\cite{SERGAN2019} that applied the relativistic GAN, and the CP-GAN\cite{CPGAN2020} which contained a densely-connected feature pyramid generator. 

\begin{table}[ht]
  \caption{Comparisons of different GAN-based SE systems. SERGAN refers the results directly adopted from the reference\cite{SERGAN2019}, while SERGAN* represents the results of those that are reimplemented using the public code\cite{SERGAN2019}. Result columns marked with '-' represents the results which cannot be reimplemented or are not provided in original papers. -SNR and -L1 represent different penalty items of the generator. The numbers with bold font are the best results among the different models.}
  \label{tab:result}
  \centering
  \scriptsize
  \begin{tabular}{ccccccc}
    \toprule
    {Model} & {PESQ} & {STOI} & {segSNR} & {Csing} & {Cbak} & {Covl} \\
    \midrule
    {SEGAN\cite{SEGAN2017}} & 2.16 & 0.925 & 7.73 & 3.48 & 2.94 & 2.80 \\
    {SERGAN\cite{SERGAN2019}} & 2.59& 0.942&-&-&-& - \\
    {SERGAN*} & 2.51&0.938& 9.36 & 3.79 & 3.24 & 3.14\\
    {CP-GAN\cite{CPGAN2020}} & 2.64 & 0.942 & - & 3.93 & 3.33 & 3.28 \\
    \midrule
    {TDCGAN-L1} & \textbf{2.87} & \textbf{0.945} & 9.82 & \textbf{4.17} & \textbf{3.46} & \textbf{3.53}\\
    {TDCGAN-SNR} & 2.79 & 0.944 & \textbf{9.97} & 4.10 & 3.43 & 3.44\\
    \bottomrule
  \end{tabular}
  \vspace{-1.5em}
\end{table}

\subsection{Results}
\label{ssec:results}

The SE performances in the context of different evaluation metrics are shown in the Table \ref{tab:result}. In the last two rows, we first compared the effect of different penalty items of generator on SE performance. We can see that the metric segSNR of TDCGAN model with SNR regularization is higher than that with L1, which implies that SNR penalty item is more helpful to improve the SNR of speech signals. Compared with these baselines systems, our proposed method outperforms them for all metrics except for the segSNR\footnote{The segSNR in the paper and the segSNR in \cite{CPGAN2020} are obtained by different calculation methods.} of CP-GAN which was not provided. The results proved that the proposed TDCGAN is more capable of removing noise from speech signals than those baselines are.

\begin{table}[ht]
  \caption{The parameter number (Millions) of different GAN-based models for SE. '$>$' is the greater-than symbol.}
  
  \label{tab:parameters}
  \centering
  \scriptsize
  \begin{tabular}{cccc}
    \toprule{}
    {SEGAN} & {SERGAN} & {CP-GAN} & {Ours} \\
    \midrule{}
    97.47 & 82.24 & $>$26.02 & \textbf{5.12} \\
    \bottomrule
  \end{tabular}
\vspace{-1.5em}
\end{table}

In order to compare the model size of different SE models, we made statistics on the number of trainable parameters of them. The results are shown in Table \ref{tab:parameters}. Because we can't reimplement CP-GAN\cite{CPGAN2020} to obtain its accurate number of parameters, we only calculated the lowest number of parameter according to the descriptions in \cite{CPGAN2020}. From the table, we can see that, when compared with the baselines SEGAN, SERGAN and CP-GAN, the number of trainable parameter of the proposed method decreased by about 19 times, 16 times and 5 times respectively. In conclusion, our TDCGAN for speech enhancement can achieve better performance  using fewer parameters than other methods do.

\section{Conclusions}
\label{sec:conclusions}

In this work, we proposed a temporal dilated convolutional generative adversarial network (TDCGAN) for speech enhancement, which further enriches the techniques of end-to-end speech enhancement. To our knowledge, it is the first time to introduce the temporal dilated convolutional network with depthwise separable convolutions and signal-to-noise ratio (SNR) gradient penalty item to the GAN architecture. For the purpose of stable training and convergence of model, we also employed some training techniques, including the simplified zero-centered gradient penalties and two-timescale-update-rule with different learning rates. The experimental results demonstrated that speech enhancement performance of the proposed method outperformed the existing state-of-the-art end-to-end GAN-based SE methods. Moreover, compared with previous methods based on GANs, the TDCGAN greatly decreases the number of trainable parameters. This will be great meaningful to push forward the applications of speech enhancement in realistic speech systems.


\bibliographystyle{IEEEbib}
\bibliography{strings,refs}

\end{document}